\documentstyle[11pt,aaspp4,flushrt]{article}

\slugcomment{\it Astron. J. {\bf 125} (2003), 2759-2768.}
\lefthead{Stern, Holden, Stanford, \& Spinrad}
\righthead{Radio-Selected Group at $z = 1.11$}

\def\cf{{c.f.,~}}
\def\ie{{i.e.,~}}
\def\eg{{e.g.,~}}

\def\mg{MG~0442+0202}
\def\cl{CL~0442+0202}

\def\deg{\ifmmode {^{\circ}}\else {$^\circ$}\fi}

\def\secper{\ifmmode \rlap.{^{s}}\else $\rlap{.}{^{s}} $\fi}

\def\kms{\ifmmode {\rm\,km\,s^{-1}}\else
    ${\rm\,km\,s^{-1}}$\fi}
\def\kmsMpc{\ifmmode {\rm\,km\,s^{-1}\,Mpc^{-1}}\else
    ${\rm\,km\,s^{-1}\,Mpc^{-1}}$\fi}
\def\ergAcm2{\ifmmode {\rm\,ergs\,cm^{-2}\,{\rm \AA}^{-1}}\else
    ${\rm\,ergs\,cm^{-2}\,\AA^{-1}}$\fi}
\def\cm2{\ifmmode {\rm\,cm^{-2}}\else
    ${\rm\,cm^{-2}}$\fi}
\def\ergcm2s{\ifmmode {\rm\,ergs\,cm^{-2}\,s^{-1}}\else
    ${\rm\,ergs\,cm^{-2}\,s^{-1}}$\fi}
\def\cgsdeg2{\ifmmode {\rm\,ergs\,cm^{-2}\,s^{-1}\,deg^{-2}}\else
    ${\rm\,ergs\,cm^{-2}\,s^{-1}\,deg^{-2}}$\fi}
\def\ergsHz{\ifmmode {\rm\,ergs\,s^{-1}\,Hz^{-1}}\else
    ${\rm\,ergs\,s^{-1}\,Hz^{-1}}$\fi}
\def\ergs{\ifmmode {\rm\,ergs\,s^{-1}}\else
    ${\rm\,ergs\,s^{-1}}$\fi}
\def\ergsA{\ifmmode {\rm\,ergs\,s^{-1}\,\AA^{-1}}\else
    ${\rm\,ergs\,s^{-1}\,\AA^{-1}}$\fi}
\def\WHz{\ifmmode {\rm\,W\,Hz^{-1}}\else
    ${\rm\,W\,Hz^{-1}}$\fi}
\def\WHzsr{\ifmmode {\rm\,W\,Hz^{-1}\,sr^{-1}}\else
    ${\rm\,W\,Hz^{-1}\,sr^{-1}}$\fi}
\def\ergscm2Hz{\ifmmode {\rm\,ergs\,cm^{-2}\,s^{-1}\,Hz^{-1}}\else
    ${\rm\,ergs\,cm^{-2}\,s^{-1}\,Hz^{-1}}$\fi}

\def\spose#1{\hbox to 0pt{#1\hss}}
\def\simlt{\mathrel{\spose{\lower 3pt\hbox{$\mathchar"218$}}
     \raise 2.0pt\hbox{$\mathchar"13C$}}}
\def\simgt{\mathrel{\spose{\lower 3pt\hbox{$\mathchar"218$}}
     \raise 2.0pt\hbox{$\mathchar"13E$}}}

\def\oii{[\ion{O}{2}] $\lambda$3727}

\def\plotfiddle#1#2#3#4#5#6#7{\centering \leavevmode
\vbox to#2{\rule{0pt}{#2}}
\includegraphics{#1}}

\begin{document}

\title{Spectroscopic Confirmation of \\
A Radio-Selected Galaxy Overdensity at $z = 1.11^1$}

\author{Daniel Stern}

\affil{JPL, California Institute of Technology, Mail Stop 169-327,
Pasadena, CA 91109 {\tt stern@zwolfkinder.jpl.nasa.gov}}

\author{Brad Holden \& S.A. Stanford}

\affil{IGPP, Lawrence Livermore National Laboratory, L-413, Livermore,
CA 94550}

\affil{Physics Department, University of California at Davis,
Davis, CA 95616 {\tt [bholden,adam]@igpp.ucllnl.org}}

\author{Hyron Spinrad}

\affil{Astronomy Department, University of California at Berkeley,
Berkeley, CA 94720 {\tt spinrad@bigz.berkeley.edu}}

\altaffiltext{1}{Some of the data presented herein were obtained at the
W.M. Keck Observatory, which is operated as a scientific partnership
among the California Institute of Technology, the University of
California and the National Aeronautics and Space Administration. The
Observatory was made possible by the generous financial support of the
W.M. Keck Foundation.  }

\begin{abstract}

We report the discovery of a galaxy overdensity at $z = 1.11$
associated with the $z = 1.110$ high-redshift radio galaxy \mg.  The
group, \cl, was found in a near-infrared survey of $z > 1$ radio
galaxies undertaken to identify spatially-coincident regions with a
high density of objects red in $I - K'$ color, typical of $z > 1$
elliptical galaxies.  Spectroscopic observations from the Keck~I
telescope reveal five galaxies within 35\arcsec\ of \mg\ at $1.10 <
z < 1.11$.  These member galaxies have broad-band colors and optical
spectra consistent with passively-evolving elliptical galaxies formed
at high redshift.  Archival {\it ROSAT} observations reveal a 3$\sigma$
detection of soft X-ray emission coincident with \cl\ at a level five
times greater than expected for the radio galaxy.  These data are
suggestive of a rich galaxy cluster and inspired a 45~ks {\it Chandra
X-Ray Observatory} observation.  As expected, the radio galaxy is
unresolved to {\it Chandra}, but is responsible for approximately half
of the observed X-ray flux.  The remaining {\it ROSAT} flux is resolved
into four point sources within 15\arcsec\ of the radio galaxy,
corresponding to a surface density two orders of magnitude higher than
average for X-ray sources at these flux levels ($S_{\rm 0.5-2 keV} > 5
\times 10^{-16} \ergcm2s$).  One of these point sources is identified
with a radio-quiet, type~II quasar at $z = 1.863$, akin to sources
recently reported in deep {\it Chandra} surveys.  The limit on an
extended hot intracluster medium in the {\it Chandra} data is $S_{\rm
1-6 keV} < 1.9 \times 10^{-15} \ergcm2s$ ($3 \sigma$,
30\arcsec\ radius aperture).  Though the X-ray observations do not
confirm the existence of a massive, bound cluster at $z > 1$, the
success of the optical/near-infrared targeting of early-type systems
near the radio galaxy validates searches using radio galaxies as
beacons for high-redshift large-scale structure.  We interpret \cl\ to
be a massive cluster in the process of formation.

\end{abstract}

\keywords{cosmology: observations -- X-rays: galaxy clusters --
galaxies:  active -- galaxies: individual (MG~0442+0202) -- galaxies:
evolution}

\section{Introduction}

The study of rich galaxy clusters at high redshift has important
consequences for our understanding of structure formation in the
Universe and is a crucial test of cosmological models.  Numerical
simulations of hierarchical models such as cold dark matter (CDM)
predict few massive clusters at large redshift \markcite{Cen:94}(\eg Cen \& Ostriker 1994),
and that the evolution of cluster number density as a function of X-ray
luminosity and temperature depends sensitively upon $\Omega_0$, but
only weakly upon $\Lambda$ and the initial power spectrum
\markcite{Peebles:89, Evrard:89, Eke:98}(\eg Peebles, Daly, \& Juszkiewicz 1989; Evrard 1989; Eke {et~al.} 1998).  Moderate-redshift
clusters from well-defined samples such as the {\it ROSAT} Deep Cluster
Survey \markcite{ROSATi:98}(RDCS; Rosati {et~al.} 1998) have been used to constrain $\Omega_M$
and $\sigma_8$ \markcite{Borgani:01}(Borgani 2001).  Distant X-ray luminous clusters
provide the best lever arm for these studies.  However, to date, few
($\approx 7$) $z > 1$ clusters have been spectroscopically confirmed
\markcite{Dickinson:95, Stanford:97, ROSATi:99, Liu:00, ROSATi:01,
Thompson:01, Stanford:02}(\cf Dickinson 1995; Stanford {et~al.} 1997; Rosati {et~al.} 1999; Liu {et~al.} 2000; Rosati 2001; Thompson {et~al.} 2001; Stanford {et~al.} 2002).

Clusters and groups of galaxies also provide a crucial tool in the
study of galaxy formation and evolution.  Out to at least $z \sim 1,$
clusters tend to be dominated by a population of massive ellipticals
which is largely homogenous and has been quiescent since at least $z
\sim 1$ \markcite{Stanford:98}(\eg Stanford, Eisenhardt, \&  Dickinson 1998).  Finding high-redshift, massive
elliptical systems is difficult, but the implications for the epoch of
early-type galaxy formation can be provocative, as evidenced by
LBDS~53W091, a galaxy at $z=1.55$ whose 3.5 Gyr age is comparable to
the Hubble time for its redshift \markcite{Dunlop:96, Spinrad:97}(Dunlop {et~al.} 1996; Spinrad {et~al.} 1997).  An
expanded census of dense environments in the early Universe will
provide a powerful means to test models of large scale structure
formation, characterize the galaxy populations in these environments,
and study the formation epoch of early-type galaxies.

Most bound clusters beyond redshift unity have been identified from
deep, serendipitous X-ray surveys, deep near-IR imaging surveys, and/or
around powerful 3C radio sources.  Radio galaxies are robust signposts
of early collapse.  In the local Universe, bright radio sources are
often hosted by giant elliptical and cD galaxies residing within dense
environments, and this identification has been shown to continue to
higher redshift --- \eg 3C184 at $z=0.996$ \markcite{Deltorn:97}(Deltorn {et~al.} 1997), 3C324 at
$z=1.21$ \markcite{Dickinson:95}(Dickinson 1995), 3C294 at $z=1.786$ \markcite{Fabian:01}(Fabian {et~al.} 2001),
and TN~J1338$-$1942 at $z = 4.1$ \markcite{Venemans:02}(Venemans {et~al.} 2002).  \markcite{Hall:01}Hall {et~al.} (2001)
also show an excess population of red galaxies around radio-loud
quasars at $1 \simlt z \simlt 2$, interpreted as being due to rich
environments associated with the quasars.

In terms of studying the intracluster gas, however, clusters around the
{\em most luminous} high-redshift radio galaxies (HzRGs) and radio-loud
quasars are less than ideal, as observed X-ray emission does not
necessarily derive solely from the hot gas.  \markcite{Hardcastle:98}Hardcastle, Lawrence, \&  Worrall (1998) show
that unresolved, soft X-ray flux correlates with 5~GHz core flux for
HzRGs and core-dominated radio-loud quasars.  {\it Chandra} has also
revealed soft, extended X-ray emission coincident with the outer radio
lobes in powerful 3C sources \markcite{Wilson:00, Harris:00}(\eg Wilson, Young, \& Shopbell 2000; Harris {et~al.} 2000).  This
emission is thought to be due to inverse-Compton scattering of cosmic
microwave background photons by the relativistic plasma in radio
jets/lobes.  At larger redshifts, the energy density of the microwave
background is higher, so that the cooling time for relativistic
electrons is short.  AGN-related soft X-ray emission should only be
problematic in the youngest and most luminous radio sources at high
redshift.  Therefore, the next obvious step in identifying a sample of
high-redshift clusters well-suited to X-ray follow-up is to locate
clusters identified with {\em less powerful} radio galaxies.

In this paper, we report the discovery of an overdensity of galaxies at
$z=1.11$ associated with the moderate-strength
\markcite{Griffiths:95}($S_{\rm 5 GHz} = 110 \pm 12$~mJy; Griffiths
{et~al.} 1995), narrow-lined HzRG \mg.  Though the X-ray observations
detailed herein do not confirm a hot intracluster medium associated
with this overdensity, the technique is sound and the radio galaxy
clearly marks a rich location in the early Universe.  The group,
CL~0442+0202, was identified from a deep optical/near-IR imaging survey
of radio sources at $z > 1$ selected from the 5~GHz MIT-Greenbank (MG)
radio catalog \markcite{Bennett:86, Stern:99a}(Bennett {et~al.} 1986;
Stern {et~al.} 1999).  \mg\ is a moderately-steep spectrum radio source
($\alpha_{\rm 1.4 GHz}^{\rm 5 GHz} = -1.05$, where $S_\nu \propto
\nu^\alpha$), approximately a quarter as luminous at 1.4~GHz compared
to the typical $z \approx 1$ 3C HzRG.  The 365~MHz morphology is
double, extended by $16\arcsec \pm 4 \arcsec$ at a position angle of
$-98\deg \pm 11\deg$ \markcite{Douglas:96}(Douglas {et~al.} 1996).  The
5~GHz morphology, measured with the VLA, reveals a triple morphology,
with two nearly equal intensity lobes separated by approximately
9\arcsec\ at a position angle of 30\deg, flanking a centrally-located
core with a flux 24\%\ of the mean lobe radio flux.  Initial optical
identification and redshift determination of \mg\ were obtained at Lick
Observatory as part of the Berkeley effort to study MG sources
\markcite{Spinrad:93, Stern:97}(\eg Spinrad {et~al.} 1993; Stern
{et~al.} 1997).

We assume $H_0 = 50~ h_{50}~ \kmsMpc$, $\Omega_M = 1$, and
$\Omega_\Lambda = 0$ throughout.  For this cosmology, the luminosity
distance of \cl\ is $7.88~ h_{50}^{-1}$~Gpc and one arcsecond subtends
$8.58~ h_{50}^{-1}$~ proper kpc.   For $H_0 = 65~ \kmsMpc$, $\Omega_M =
0.35$, and $\Omega_\Lambda = 0.65$, the cosmology favored by recent
high-redshift supernovae and cosmic microwave background observations,
these distances change very slightly:  they are smaller by 0.4\%.
Throughout we have corrected for foreground Galactic extinction using a
reddening of $E_{B-V} = 0.155$ determined from the dust maps of
\markcite{Schlegel:98}Schlegel, Finkbeiner, \& Davis (1998).

\section{Observations and Results}

\subsection{Optical and near-infrared imaging}

To search for red galaxies associated with the HzRG \mg, we obtained
deep, near-infrared imaging of the field using the Gemini Twin-Arrays
Infrared Camera \markcite{McLean:94}(McLean {et~al.} 1994) on the Lick 3~m Shane telescope.
Gemini uses a dichroic beam splitter which yields two independent
beams:  a short-wave channel with a Rockwell NICMOS~3 $256 \times 256$
HgCdTe array and a long-wave channel with a SBRC $256 \times 256$ InSb
array.  Gemini was used in its 0\farcs7 pix$^{-1}$ mode, providing a
3\arcmin\ $\times$ 3\arcmin\ field of view.  Our data, obtained through
$J$ ($\lambda_{\rm c} = 1.25~ \mu$m; $\Delta \lambda = 0.29~ \mu$m) and
$K'$ ($\lambda_{\rm c} = 2.12~ \mu$m; $\Delta \lambda = 0.34~ \mu$m)
filters for the two channels respectively, were obtained in
non-photometric conditions on UT 1999 January 3 and UT 1999 December
2.  The total integration time was 12~ks, and the combined $K'$ image,
reporting 2\farcs3 seeing, is provided in Fig.~\ref{fig_Kimage}.

To calibrate these images, we obtained $JK'$ snapshots of the
\mg\ field on UT 1999 February 26 in photometric conditions and
0\farcs7 seeing with NSFCAM \markcite{Shure:94}(Shure {et~al.} 1994) at the 3~m Infrared
Telescope Facility (IRTF).  NSFCAM was used in its 0\farcs3 pix$^{-1}$
mode, providing a 77\arcsec\ $\times$ 77\arcsec\ field.  Observations
of standard stars from the \markcite{Persson:98}Persson {et~al.} (1998) list were used to
calibrate the images onto the California Institute of Technology (CIT)
system, defined in \markcite{Elias:82}Elias {et~al.} (1982).  The calibrated Lick Observatory
$K'$ image reaches a depth of $K' = 20.0$ (3$\sigma$ in a 4\farcs6
diameter aperture).

We obtained an optical, $I$-band ($\lambda_{\rm eff} =$ 8331 \AA;
$\Delta \lambda$ = 3131 \AA) image of \mg\ on UT 2001 January 28 with
the Low Resolution Imaging Spectrometer \markcite{Oke:95}(LRIS; Oke {et~al.} 1995) on the
Keck~I telescope.  This 600~s exposure, calibrated with observations of
PG~1323$-$086 \markcite{Landolt:92}(Landolt 1992) obtained the same night, has 0\farcs74
seeing and reaches a depth of 24.7 mag ($3 \sigma$ in 1\farcs5 diameter
aperture).  LRIS uses a $2048 \times 2048$ backside-illuminated Tek CCD
with 0\farcs212 pix$^{-1}$, providing a 6\arcmin\ $\times$ 7\farcm8
field of view.

After geometrically transforming the images to a common frame, we used
SExtractor \markcite{Bertin:96}(release V2.1.6; Bertin \& Arnouts 1996) to create catalogs of
sources selected from the $K'$ images.  Photometry was generated for
6\arcsec\ diameter apertures.  Fig.~\ref{fig_colormag} presents
resultant color-magnitude diagrams for the Lick and IRTF $K'$ images.
Compared to random extragalactic fields, the field of \mg\ has a high
surface density of objects with red $I - K'$ colors.  Using the IRTF
infrared imaging, we find six sources in the 1.6 arcmin$^2$ field with
$I-K > 4$ and $K < 20$, corresponding to a surface density of extremely
red objects (EROs) of 3.8 EROs arcmin$^{-2}$.  This is 17 times the
surface density of such sources found by \markcite{Barger:99b}Barger {et~al.} (1999) in a 61.8
arcmin$^2$ survey and $\approx 8$ times the surface density of such
sources found in the 47.2 arcmin$^2$ Herschel Deep Field
\markcite{McCracken:00}(McCracken {et~al.} 2000).  The dotted line in Fig.~\ref{fig_colormag} at $I
- K' \sim 4.5$ shows the red sequence characteristic of local cluster
galaxies, as determined by band-shifting photometry of Coma cluster
galaxies \markcite{Stanford:98}(see Stanford {et~al.} 1998).  Particularly in the better-seeing
IRTF imaging where confusion is less troublesome, photometric errors
less severe, and the smaller field less contaminated by field galaxies,
a red sequence is evident approximately 0.5~mag blue-ward of the
no-evolution prediction.  This amount of bluing in observed $I - K'$
color is consistent with the color change due to passive evolution of a
single-age, solar-metallicity, stellar population at $z = 1.11$ formed
in a 0.1~Gyr burst at a formation redshift $z_f = 3$ \markcite{Bruzual:93}(using the
updated 2000 GISSEL models of Bruzual \& Charlot 1993).  We targeted this red
sequence for spectroscopic investigation.

\subsection{Keck spectroscopy}

We obtained optical spectra of photometric candidates using LRIS in
slitmask mode on the Keck~I telescope.  Masks on UT 2001 January 30 and
UT 2001 February 19 were observed for 1.5~hr, while a fainter mask on
UT 2002 February 5$-$6 was observed for 3.5~hr.  On UT 2002 March 10 we
observed a mask mainly targeting {\it Chandra} sources in the field for
1.5~hr.   A shallow, 20~min longslit observation of two bright {\it
Chandra} sources was also obtained on UT 2002 March 11 during
twilight.  All observations used the 400 lines mm$^{-1}$ grating
($\lambda_{\rm blaze} = 8500$ \AA; spectral resolution $\Delta
\lambda_{\rm FWHM} \approx 8$ \AA) and sample the wavelength range
$\approx 5000$ \AA\ to 1 $\mu$m.  The March 2002 observations also
implemented the blue side of LRIS, using a D560 dichroic and the 300
lines mm$^{-1}$ grism ($\lambda_{\rm blaze} = 5000$ \AA; spectral
resolution $\Delta \lambda_{\rm FWHM} \approx 13.5$ \AA), providing
spectral coverage down to $\approx 3500$ \AA.  The observations were
split into 1800~s exposures with $\approx$ 3\arcsec\ spatial offsets
performed between integrations to facilitate removal of fringing at
long wavelengths.  Data reductions followed standard procedures and the
spectra were flux-calibrated using observations of standard stars from
\markcite{Massey:90}Massey \& Gronwall (1990).  Sample optical spectra are presented in
Fig.~\ref{fig_spec} and results for all $K'$-detected sources are
presented in Table~\ref{tableProp}.  Redshifts were determined by
visual inspection of emission and absorption features.  A histogram of
the redshifts we have obtained thus far in the \mg\ field is presented
in Fig.~\ref{fig_zhist}, showing a pronounced redshift spike at $z \sim
1.1$.  Besides \mg, a total of seven color-selected targets within
$\simlt$ 30\arcsec\ of the radio galaxy were attempted
spectroscopically, of which five are confirmed cluster members and one
has an inconclusive spectrum.  Most of the confirmed cluster members
have spectra typical of local early-type galaxies, showing strong 4000
\AA\ breaks and weak or no emission lines.  The exception is, of
course, the HzRG \mg\ which shows high equivalent width, narrow, high
ionization state emission features, typical of the HzRG population
\markcite{McCarthy:93}(\eg McCarthy 1993).

\subsection{{\it ROSAT} archive }

A search of the {\it ROSAT} archive reveals that High Resolution Imager
\markcite{Trumper:83}(HRI; Tr\"umper 1983) observations of the $z=0.19$ cluster
MS~0440+0204 cover CL 0442+0202 at an off-axis angle of 13\farcm4.
These data, discussed in \markcite{Gioia:98}Gioia {et~al.} (1998), were obtained in 1994
February/March and 1995 August/September.  We processed the archival
data, which total 27.2~ks, and identify a 3.3$\sigma$ X-ray source
slightly (2$\sigma$ of the point spread function) offset from the radio
galaxy.  The resolution of {\it ROSAT} at the large off-axis angle of
CL~0442+0202 is inadequate for determining if the X-ray emission is
extended or point-like.  After correcting for the expected soft X-ray
emission from the radio galaxy \markcite{Hardcastle:98}(Hardcastle {et~al.} 1998), the {\it ROSAT}
detection suggested soft X-ray emission of $\approx 10^{-13} \ergcm2s$,
consistent with a high-redshift hot intracluster medium (ICM).  This
exciting hypothesis inspired an observation with the {\it Chandra X-Ray
Observatory} \markcite{Weisskopf:96}(Weisskopf, {O'dell}, \&  {van~Speybroeck} 1996).

\subsection{{\it Chandra} observations}

On UT 2002 February 10 we obtained a 44.1~ks observation of \cl\ with
{\it Chandra}.  Compared to the {\it ROSAT} image, these data (OBS-ID
03242), taken with the Advanced CCD Imaging Spectrometer (ACIS-I),
reach a greater depth with significantly improved ($\approx 0\farcs5$)
angular resolution.  We reduced and analyzed the data following
standard procedures, using the {\it Chandra} Interactive Analysis of
Observations (CIAO) software (release V2.1.2).  Sources were identified
independently in both soft (0.5$-$2~keV; $S$) and hard (2$-$7~keV; $H$)
energy bands.  Following previous analyses of deep {\it Chandra} data
\markcite{Tozzi:01, Stern:02b}(\eg Tozzi {et~al.} 2001; Stern {et~al.} 2002b), we determined a hardness ratio $(H
- S)/(H + S)$, using the background-subtracted counts.  A response
matrix and effective area were then determined for each detected
source.  Assuming a Galactic $N_{\it H} = 9.49 \times 10^{20} {\rm
cm}^{-2}$ and a typical photon index, $\Gamma = 1.4$, counts in each
band were converted into fluxes.  Table~2 presents the X-ray results
for those {\it Chandra} sources contained within the optical image,
along with optical magnitudes and spectroscopic redshifts, when
available.

\section{Discussion}

\subsection{{\it Chandra} results}

The most striking result of the {\it Chandra} data is that the improved
spatial resolution resolves the {\it ROSAT} detection into five point
sources.  The total soft-band flux of these sources accounts for only
20\%\ of the {\it ROSAT} detection, likely due to the low significance
of the {\it ROSAT} detection with temporal variability also possible.
As expected, the radio galaxy \mg\ is detected with a relatively hard
X-ray spectrum; we discuss this source further in \S 3.2.  At the depth
of our image, $S_{\rm 0.5-2 keV} \approx 5 \times 10^{-16} \ergs$, {\it
Chandra} observations of deep fields find a mean surface density of
approximately 1000 sources deg$^{-2}$ \markcite{Stern:02b}(\eg Stern {et~al.} 2002b).  The
five sources we detect in our {\it Chandra} image, all within
15\arcsec\ of the HzRG, represent a surface density approximately two
orders of magnitude higher!  Unfortunately, this unusual configuration
of point sources rather than an extended ICM is likely responsible for
the {\it ROSAT} detection.  We find no diffuse emission associated with
\cl\ (see \S 3.4).

Of the twelve {\it Chandra} sources within the optical image, we have
unambiguous spectroscopic redshifts for six and and tentative redshifts
for an additional two.  Typical of follow-up of {\it Chandra} sources,
the two sources with the softest ($H-S/H+S \approx -0.5$) X-ray spectra
are identified with luminous, broad-lined quasars.  Both have projected
celestial positions close to \cl\ (see Table~1), but one (CXO~6) is
background while the other (CXO~7) is foreground.  Potentially they could
provide useful probes of the environment of \cl.  The two sources with
ambiguous spectra both have relatively soft X-ray spectra and optical
spectra that reveal blue continuum, each with the detection of a single,
broad emission line.  We tentatively interpret both lines as \ion{Mg}{2}
$\lambda$ 2800\AA: at $z \approx 1$, the strong rest-frame UV lines
characteristic of quasars have not yet entered the optical window, while
the strong rest-frame optical lines have shifted to near-IR wavelengths,
either beyond the optical window or at redder wavelengths where the sky
is less benign.

The source CXO~10 is typical of the new population of sources being
identified in deep {\it Chandra} images \markcite{Hornschemeier:01,
Stern:02b}(\eg Hornschemeier {et~al.} 2001; Stern {et~al.} 2002b):  galaxies whose optical/near-IR properties suggest a normal
galaxy with no indication of an active nucleus, but are nevertheless
luminous X-ray sources.  Many of these sources, including CXO~10, have
hard X-ray spectra, implying significant obscuration of a buried
central engine.  In the case of CXO~10, the optical spectrum is
dominated by an old stellar population with a strong 4000~\AA\ break.
Weak [\ion{O}{2}] $\lambda$ 3727\AA\ is also present, possibly due to
low levels of star formation, possibly due to the narrow-line region of
the buried, active nucleus.

Finally, our {\it Chandra} data reveal two obvious AGN showing high
equivalent width, high ionization state, narrow emission lines, but
lacking the obvious broad-line signatures associated with quasars:
CXO~5 and CXO~12.  Following Seyfert nomenclature, these sources are
often called {\em type II quasars}, and deep X-ray data has sparked a
renewed interest in the population \markcite{Norman:02, Stern:02a,
Dawson:03}(\eg Norman {et~al.} 2002; Stern {et~al.} 2002a; Dawson {et~al.} 2003).  The radio-loud end of this population has been studied for
several decades:  indeed, CXO~5 is identified with \mg\ and we discuss
it in more detail in \S 3.2.  CXO~12 is the other type~II active system
identified in our data.  The X-ray spectrum is relatively flat from the
soft to the hard bands ($H-S/H+S = -0.03$), providing a harder X-ray
spectrum than the typical, bright X-ray population.  The optical
spectrum (Fig.~\ref{fig_qso2}) shows narrow, rest-frame UV emission
lines with very little continuum.  \markcite{Stern:02a}Stern {et~al.} (2002a) discusses a
similar source in detail \markcite{Norman:02, Dawson:03}(see also Norman {et~al.} 2002; Dawson {et~al.} 2003).
Both the apparently-normal galaxies and these type~II quasars are being
found in large numbers in surveys with the new generation of X-ray
satellites.  Such studies will be essential for creating an unbiased
census of AGN in the Universe, thereby testing models of the X-ray
background and providing a history of accretion-driven energy
production in the Universe.

\subsection{The radio galaxy MG~0442+0202}

The $z = 1.11$ radio galaxy \mg\ is what initially drew our attention
to this field.  With $K' = 17.1$, \mg\ fits smoothly onto the
surprisingly low-scatter Hubble, or $K-z$, relation of radio galaxies
\markcite{DeBreuck:01}(\eg see Fig.~10 in {De~Breuck} {et~al.} 2001).  At each redshift, HzRGs are
the most luminous galaxies known at observed 2$\mu$m.  Since rest-frame
2$\mu$m samples stellar emission from the low-mass stars which dominate
the baryon content of a galaxy, this is generally interpreted as HzRGs
being the most massive systems at each cosmic epoch.  Indeed, our deep,
optical spectrum of \mg\ (Fig.~\ref{fig_spec}) shows that this $z =
1.11$ radio source is hosted by an old, massive galaxy.  Though the
dominant spectral features are strong, narrow emission lines
characteristic of radio galaxies \markcite{McCarthy:93}(\eg McCarthy 1993), the
spectrum also clearly shows stellar light.  Absorption is seen from
Ca~H $\lambda$ 3933\AA\ and a conglomerate of \ion{Fe}{1}/\ion{Mg}{1}
lines around 3840 \AA\ \markcite{Pickles:85}(\eg Pickles 1985), indicative of
low-mass stars.  Ca~K $\lambda$ 3968.5 \AA\ has been filled in by
[NeIII] $\lambda$ 3967.5 \AA\ emission.  The strong 4000~\AA\ break
seen in \mg\ is another strong indicator of stellar emission in this
HzRG.  The break is due to the sudden onset of stellar photospheric
opacity shortward of 4000 \AA, associated with the Fraunhofer H and K
lines of \ion{Ca}{2}, as well as various ionization stages of other
elements heavier than helium \markcite{Ohman:34}(\"Ohman 1934).

The soft X-ray properties of core-dominated quasars are strongly
correlated with their radio properties, with a radio (5~GHz) core to
soft X-ray core two-point spectral index of $\alpha_{xr} \sim 0.85$
\markcite{Worrall:94, Worrall:97}(Worrall {et~al.} 1994; Worrall 1997).  Assuming an X-ray spectral index of
$\alpha = -0.25$ (see below), the observed flux density of \mg\ at
1~keV is 3.3~nJy.  For its 5~GHz radio-core flux density of 12~mJy,
\mg\ is slightly overluminous in the soft X-ray compared to the
relation plotted by \markcite{Worrall:94}Worrall {et~al.} (1994) for core-dominated quasars.
This result is consistent with other HzRGs examined by
\markcite{Hardcastle:98}Hardcastle {et~al.} (1998) and may indicate non-jet-related X-ray emission
originating close to the AGN.

Our {\it Chandra} observation detects more than 250 counts from \mg,
providing a robust hardness ratio of $(H - S)/(H + S) = -0.29$.
Assuming the X-ray flux spectrum can be expressed as simple power-law
function of energy $E$ [\ie $F(E) \propto E^\alpha$] modified by the
Galactic absorption column density in the direction of the source
($N_{\rm H} = 9.49 \times 10^{20}$ cm$^{-2}$), this hardness ratio
corresponds to a spectral index of $\alpha \approx -0.25$ --- slightly
flatter than the steep $\alpha \approx -0.8$ spectra that are commonly
observed for unabsorbed AGNs \markcite{Nandra:94}(\eg Nandra \& Pounds 1994).  Given the
narrow-line optical/UV spectrum of the source, it is likely that
\mg\ actually has an intrinsically steep X-ray spectrum that appears to
be flat due to significant soft X-ray absorption by intervening
material within the galaxy.

To investigate this possibility, we examined the X-ray spectrum of
\mg\ more closely.  Using the CIAO software, we extracted source counts
within a 3\farcs0 radius centered on \mg; the background was determined
from a rectangular region of 100 arcseconds by 180 arcseconds.  This
region was chosen to overlap with the ACIS-I node in which the radio
galaxy falls, node two for chip three.  All detected sources in this
region were removed.  As we were expecting a cluster in this
observation, we observed with a $-8$ mm SIM-Z offset, hence the target
did not fall on the standard aimpoint.  We also included a
2\arcmin\ offset in the y-direction to move the potential cluster off
of the chip gap in the ACIS-I mosaic, which moved the target
2\arcmin\ away from the optical axis.

We binned the spectrum into 26 channels of 10 events from 0.8 to 6.0
keV and we modeled the spectrum using the XSPEC v11.1.2 package
\markcite{Arnaud:96}(Arnaud 1996) with the recently-created Chartas \& Getman ACIS
absorption model\footnote{Available at {\tt
http://www.astro.psu.edu/users/chartas/xcontdir/xcont.html}.} and using
the \markcite{Markevitch:01}Markevitch \& Vikhlinin (2001) correction to the effective area of ACIS-I.
We fit a power law with an absorption of $9.49 \times 10^{20} {\rm
cm^{-2}}$ and found the best fitting slope for a simple power-law X-ray
spectrum to be $\alpha = -0.53 \pm 0.13$.  If we add an additional
absorber component at the redshift of the radio galaxy and fix the
intrinsic value of $\alpha = -0.8$, we find an column density of
$3.0^{+5.3}_{-3.0} \times 10^{21}~ {\rm cm^{-2}}$ (90\% confidence
limits).  This model implies an unabsorbed luminosity of $2.7 \pm 0.5
\times 10^{44}~ {\rm erg\ s^{-1} }$ (90\% confidence limits; 2-10 keV),
which is well within the quasar regime.


We also searched the extracted X-ray spectrum of \mg\ for redshifted
Fe\,K$\alpha$~6.4~keV emission, which would appear at observed
3.03~keV.  Fig.~\ref{fig_Xspec} presents the background-subtracted
X-ray spectrum with a best-fit continuum model discussed above,
assuming no intrinsic absorption.  We find no excess emission evident
at 3~keV, though we note that {\it Chandra}'s sensitivity falls at
these higher energies.

\subsection{Cluster galaxy members}

The color-selected galaxy overdensity associated with \mg\ affords us
the opportunity to study early-type galaxy populations at high
redshift.  As seen in Fig.~\ref{fig_spec}, many of the galaxies
associated with \mg\ have evolved stellar populations as evidenced, for
example, by strong continuum decrements at 4000~\AA.  This feature is
of particular importance as the 4000~\AA\ decrement intensifies with
the age of a stellar population, potentially providing a sensitive
probe of the age of a galaxy.  Most of the other $z > 1$ early-type
galaxy overdensities known (see \S 1) are at sufficiently high redshift
($z \simgt 1.25$) that this feature has shifted to challenging
wavelengths for silicon-based detectors, especially from the ground
where telluric emission is severe.  In such cases, astronomers are
forced into the difficult situation of age-dating galaxies based on
breaks at 2640~\AA\ and 2900 \AA, where the galaxy is several times
fainter \markcite{Spinrad:97, Yi:99}(\eg Spinrad {et~al.} 1997; Yi {et~al.} 1999).  The redshift of \mg\ is
ideal for probing the formation history of early-type galaxies:  $z =
1.11$ is sufficiently large to be cosmologically interesting, but not
so large as to be challenging spectroscopically.

$D(4000) = 1.72 - 1.85$ for the three galaxies presented in
Fig.~\ref{fig_spec}.  Conservatively assuming that the stellar
populations were formed from a single, instantaneous, solar-metallicity
burst, these galaxies have minimum ages of approximately 1~Gyr, where
we have compared the amplitudes of the D(4000) break to
\markcite{Bruzual:93}Bruzual \& Charlot (1993) stellar population synthesis models for both
\markcite{Salpeter:55}Salpeter (1955) and \markcite{Scalo:86}Scalo (1986) initial mass functions and a
range of star formation histories (Fig.~\ref{fig_d4000}).  The implied
formation redshift is $z_f \simgt 1.5$.  We note that several of the
group members show [\ion{O}{2}] emission at some level, suggesting some
ongoing star formation.  This would imply younger, bluer stars diluting
the D(4000) amplitude, and thus an earlier formation redshift.  For the
exponentially-decreasing star formation history plotted, the minimum
age is approximately 4~Gyr, requiring formation redshifts $z_f > 5$.

We now compare the absolute magnitudes of the \cl\ galaxies to local
cluster luminosity functions:  for a 1~Gyr old single-burst stellar
population, calculated using the \markcite{Worthey:94}Worthey (1994) models for a
\markcite{Salpeter:55}Salpeter (1955) initial mass function, solar metallicity, and a
mass range of $0.21 - 10 M_\odot$, the $k$-correction from observed
$K'$ to rest-frame Gunn~$i$ is 0.21~mag, implying $M_i = -24.9$ for
\mg\ for our adopted cosmology.  From a sample of 39 Abell clusters,
\markcite{Paolillo:01}Paolillo {et~al.} (2001) find $M_i^* \sim -22.4$, using the same cosmology.
During the 8.8~Gyr that elapses from $z = 1.11$ to $z = 0$, our model
predicts $\sim 1.4$~mag fading due to luminosity evolution, suggesting
that \mg\ will evolve into a $\sim 3 L^*$ elliptical in the present
epoch if there is no additional merging.  The next brightest confirmed
cluster member, galaxy\#2, has $M_i = -23.4$, suggesting a $\sim 0.7
L^*$ elliptical in the present epoch for the same assumptions.

Including \mg, we have identified six galaxies within an angular region
of 1\arcmin\ diameter that have similar velocities.  Their redshifts
have a mean of 1.104.  Jackknife determinations of the velocity
dispersion using both the gapper method and biweights yield $\sigma
\approx 600 \pm 100 \kms$.  We find similar results for the bootstrap
method.

\subsection{A Hot Intracluster Medium?}

We analyzed the {\it Chandra} data to study possible diffuse soft X-ray
emission from a hot intracluster medium.  Excluding the obvious five
point sources, we implemented a simple curve of growth analysis.  We
made a series of annuli of increasing width and subtracted off a
background annulus from 40\arcsec\ to 60\arcsec\ in radius, all
centered on \mg.  In no annulus did we achieve a $3 \sigma$ detection
of excess events in the 0.3 to 10.0 keV energy band.  In the 1.0 to 6.0
keV band, we find a 3$\sigma$ upper limit of $1.9 \times 10^{-15}
\ergcm2s$ for a 3 keV cluster at $z=1.110$ within a 30\arcsec\ radius
centered on the radio galaxy.  This radius corresponds to a 257
$h_{50}^{-1}$ kpc at the redshift of the cluster, a typical core
radius for a massive cluster of galaxies.  We choose 3 keV as the
cluster temperature based on the $\sim\ 600~ {\rm km\ s^{-1}}$ velocity
dispersion measured in \S 3.3 and equation 13 in \markcite{Wu:99}Wu, Fabiand, \& Fang (1999).
Translating the flux limit into a luminosity, we find an upper limit of
$1.5\times 10^{43}~ {\rm erg\ s^{-1} }$ for the bolometric luminosity of
the cluster.  Using the bolometric luminosity -- velocity dispersion
relation of equation 7 in \markcite{Wu:99}Wu {et~al.} (1999), we would expect a bolometric
luminosity of $2.3\times 10^{44}~ {\rm erg\ s^{-1} }$.  This would
translate into a $15 \sigma$ detection within the 30\arcsec\ radius,
assuming a core radius of 250 $h_{50}^{-1}$ kpc and a $\beta$ model
with a slope of $\frac{2}{3}$.

\section{Conclusions}

Optical/near-IR imaging of the moderate-strength, $z = 1.11$ radio
source \mg\ has revealed an overdensity of sources with red optical to
near-IR colors, typical of $z \simgt 1$ early-type galaxies.
Spectroscopic investigations with the Keck telescope have secured
redshifts for five of these galaxies in the range $z = 1.105 \pm 0.005$
within 35\arcsec\ of \mg.  These data indicate the presence of an
overdensity at high redshift.  The member galaxies form a homogenous
population with similar colors.  The colors and tightness of the
color-magnitude relation support the conclusions of \markcite{Stanford:98}Stanford {et~al.} (1998)
and \markcite{vanDokkum:01b}van Dokkum \& Franx (2001) who found that the spectrophotometric
properties of early-type cluster galaxies are consistent with the
passive evolution of an old stellar population formed at an early
cosmic epoch.  Our data (Fig.~\ref{fig_colormag}) suggests a less steep
color-magnitude relation than seen in the nearby Coma cluster.  This
effect is likely due to a blue AGN contribution to the brightest galaxy
in our group, \mg.  Similar flattenings of the color-magnitude relation
are expected near the formation epoch of cluster ellipticals, though
galactic winds in single-age monolithic models can also produce such
effects  \markcite{vanDokkum:01}(\eg van Dokkum {et~al.} 2001).

Archival {\it ROSAT} data found soft X-ray emission coincident with
this structure with an intensity higher than expected from the radio
galaxy alone.  Deeper, higher-resolution {\it Chandra} observations
reveal no extended soft X-ray emission from a hot ICM associated with
the putative cluster, instead finding 5 X-ray point sources within
15\arcsec\ of the radio galaxy, approximately 100 times the average
surface density of X-ray sources at this flux level.  The lack of
extended X-ray emission coupled with the spectroscopically-confirmed $z
= 1.11$ galaxies with a tight color-magnitude relation lead us to
interpret \cl\ to be a protocluster in the process of formation.  Based
on the sparse data presented here, and the optical faintness of the
associated galaxies, it is difficult to fully determine the nature of
the overdensity associated with \mg.  Deeper data and more extensive
spectroscopy are necessary to determine if \cl\ will become a modest
group or a rich cluster.  We can safely say, however, that \cl\ is
likely a structure which has broken away from the Hubble flow, but is
still a long way from virialization.  A weak ICM is probably already in
place, with some metal enrichment from previous generations of stars,
but it remains undetected since \cl\ is not virialized and the (future)
ICM's temperature and density are far too low to produce detectable
X-ray emission.  As noted previously \markcite{Lubin:02}(\eg Lubin, Oke, \& Postman 2002), many
optically-selected clusters at $z \simgt 0.5$ are also underluminous in
the X-rays when compared to the X-ray luminosity -- velocity ($L_X -
\sigma$) relation for local and moderate-redshift clusters.  This
suggests that clusters are actively forming at these redshifts.

Accepting that \cl\ is a cluster in the process of formation, we note
that the old age of the \cl\ member galaxies (\S 3.3) implies that the
stars in clusters are older than the ICM.  Similarly, based on {\it
Hubble Space Telescope} imaging of spectroscopically-confirmed members
of the optically-selected, X-ray-underluminous cluster CL~1324+3011 ($z
= 0.76$), \markcite{Lubin:02}Lubin {et~al.} (2002) find a high early-type fraction
($0.55^{+0.17}_{-0.14}$), indicative of a substantial tenure for these
constituent galaxies.  Since the ICM is processed gas, these results:
(1) mean the ICM is relatively young, and (2) raise the question of
which galaxies are the source of the metal enrichment.

This work shows that high-redshift radio galaxies can be beacons of
rich regions in the early Universe, allowing the identification of
clusters and protoclusters at $z \simgt 1$.  A census of galaxy
clusters at $1 \simlt z \simlt 1.5$ would provide a major breakthrough
toward our understanding of early-type galaxy formation, cluster
formation, as well as provide important constraints on basic
cosmological parameters.  Additionally, unified models of AGN predict
that HzRGs and radio-loud quasars differ only in the orientation of
their central jets relative to the observer; on average, they are
expected to reside in similar galactic environments.  This hypothesis
could be tested with a systematic study of a well-chosen sample of
radio-loud galaxies and quasars.

\acknowledgments

The authors wish to recognize and acknowledge the very significant
cultural role and reverence that the summit of Mauna Kea has always had
within the indigenous Hawaiian Community.  We are most fortunate to
have the opportunity to conduct observations from this mountain.  We
thank Leslie Maxfield for contributions to preliminary optical
observations of the radio galaxy MG~0442+0202 from Lick Observatory and
we gratefully acknowledge Charles Lawrence for enlightening discussion
regarding the X-ray properties of HzRGs.  We thank Megan Eckart and
Fiona Harrison for Keck spectroscopy of {\it Chandra} sources in this
field, obtained as part of the SEXSI survey (Harrison et al., in
preparation), and we thank Marc Davis, Alison Coil, and Ed Moran for
providing the Keck $I$-band image.  We are indebted to Ian McLean and
Elinor Gates for building and supporting the Gemini camera at Lick
Observatory, and to Kurt Adelberger for an insightful and careful
referee report.  Support for this project came from {\it Chandra} grant
GO2-3194.  The work of DS was carried out at the Jet Propulsion
Laboratory, California Institute of Technology, under a contract with
NASA.  The work by SAS and BH at Lawrence Livermore National Laboratory
was performed under the auspices of the Department of Energy under
contract W-7405-ENG-48.


\begin{figure}[!t]
\begin{center}
\end{center}

\caption{{\bf Right:}  Near-infrared Lick/Gemini $K'$ image of the $z =
1.11$ galaxy group associated with the radio galaxy \mg.  Exposure time
is 12~ks and the image is 3\farcm2 on a side.  North is up, and east
is to the left; coordinate axes provide offsets relative to \mg,
located at R.A. = 04$^h$42$^m$23.74$^s$, Dec. =
+02\deg02\arcmin19\farcs8 (J2000).  {\bf Left:} Smaller field-of-view
images from Keck/LRIS ($I$) and IRTF/NSFCAM ($K'$), 1\farcm2 on a
side.  In all images, spectroscopically-confirmed group members ($1.10
< z < 1.11$) are circled, sources with $I - K' > 3$ and $K' < 20$
spectroscopically shown not to reside in the group are boxed, while
candidate group members, selected to have $I - K' > 3$ and $K' < 20$,
are both circled and boxed.  The {\it ROSAT} HRI detection is marked
with a large ``X'' in the Lick image.}

\label{fig_Kimage}
\end{figure}

\begin{figure}[!t]
\begin{center}
\plotfiddle{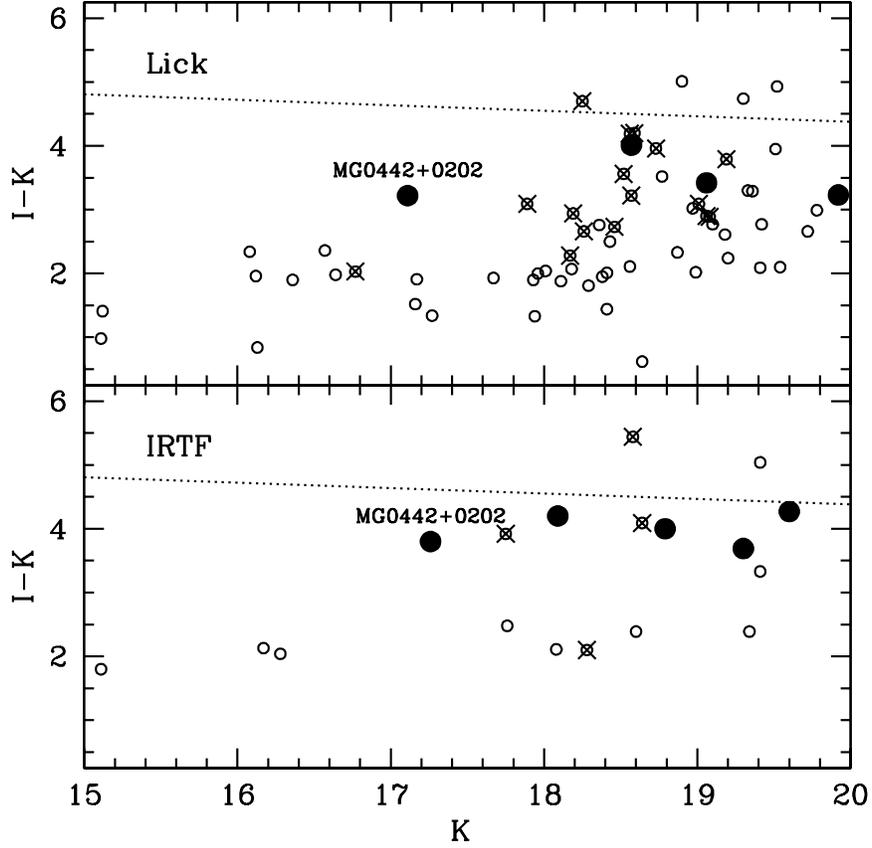}{4.2in}{0}{60}{60}{-185}{-100}
\end{center}

\caption{Color-magnitude diagrams for the field of \mg.  Top panel uses
the infrared $K'$ image from Lick Observatory, surveying 9.9
arcmin$^2$.  Bottom panel uses the infrared $K'$ image from the IRTF,
surveying 1.6 arcmin$^2$.  Photometry was generated for 6\arcsec\
diameter apertures.  Galaxies spectroscopically confirmed to reside at
$z \sim 1.11$ are marked with larger, filled symbols; \mg\ (labeled) is
the brightest of these sources.  Sources spectroscopically shown not to
reside at the redshift of \mg\ are marked with crosses.  The dotted
line represents the no-evolution prediction for early-type galaxies
based on the Coma cluster (see text).}

\label{fig_colormag}
\end{figure}

\begin{figure}[!t]
\begin{center}
\plotfiddle{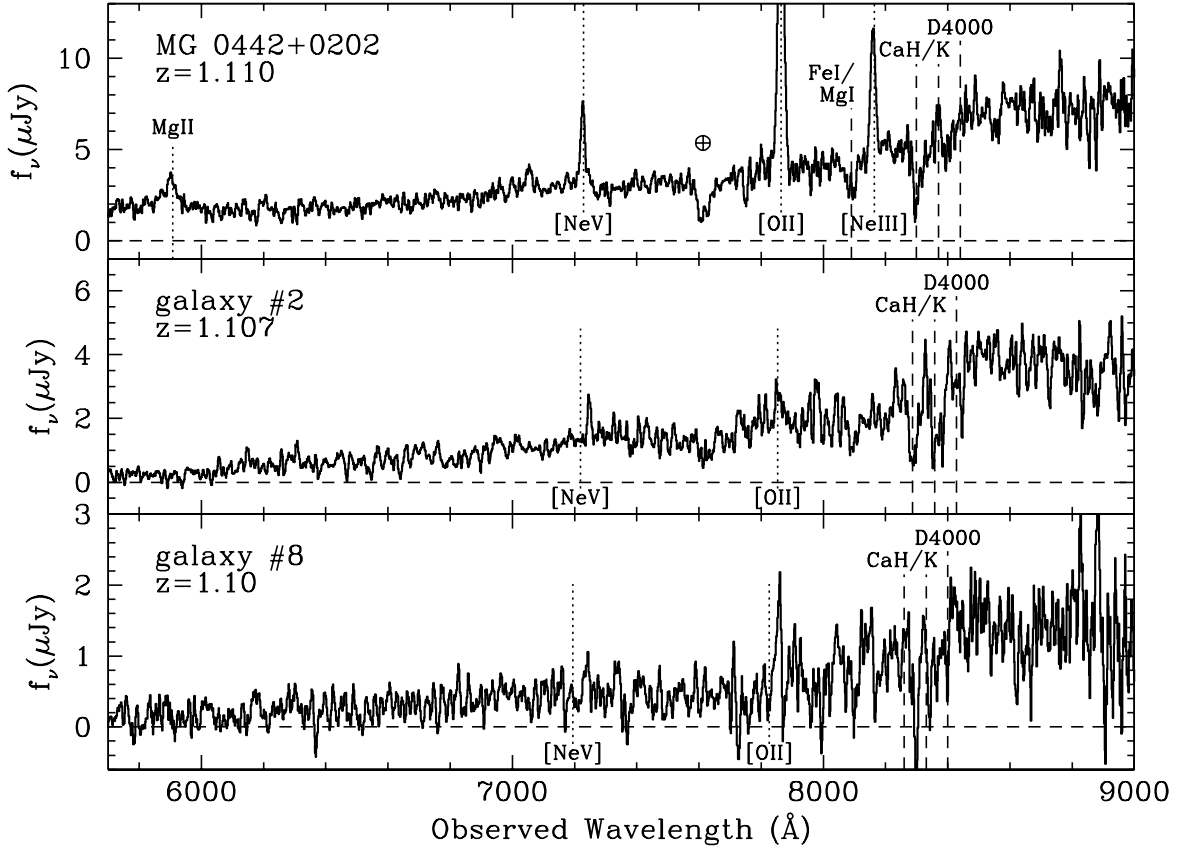}{4.0in}{-90}{60}{60}{-235}{335}
\end{center}

\caption{Optical spectra of three galaxies in \cl, obtained with the
Keck~I telescope.  Top spectrum is the high-redshift radio galaxy \mg,
showing several prominent, narrow emission lines.  Exposure times are
5.5~ks.  Spectra were extracted using a 1\farcs5 $\times$ 1\farcs5
aperture and smoothed with a 15 \AA\ boxcar filter.  Prominent features
are labeled; not all are detected.  There are hints of weak
[\ion{O}{2}] in galaxy \#2.  The feature at 7857 \AA\ in galaxy \#8
appears to be the residual of a bright night sky line.}

\label{fig_spec}
\end{figure}

\begin{figure}[!t]
\begin{center}
\plotfiddle{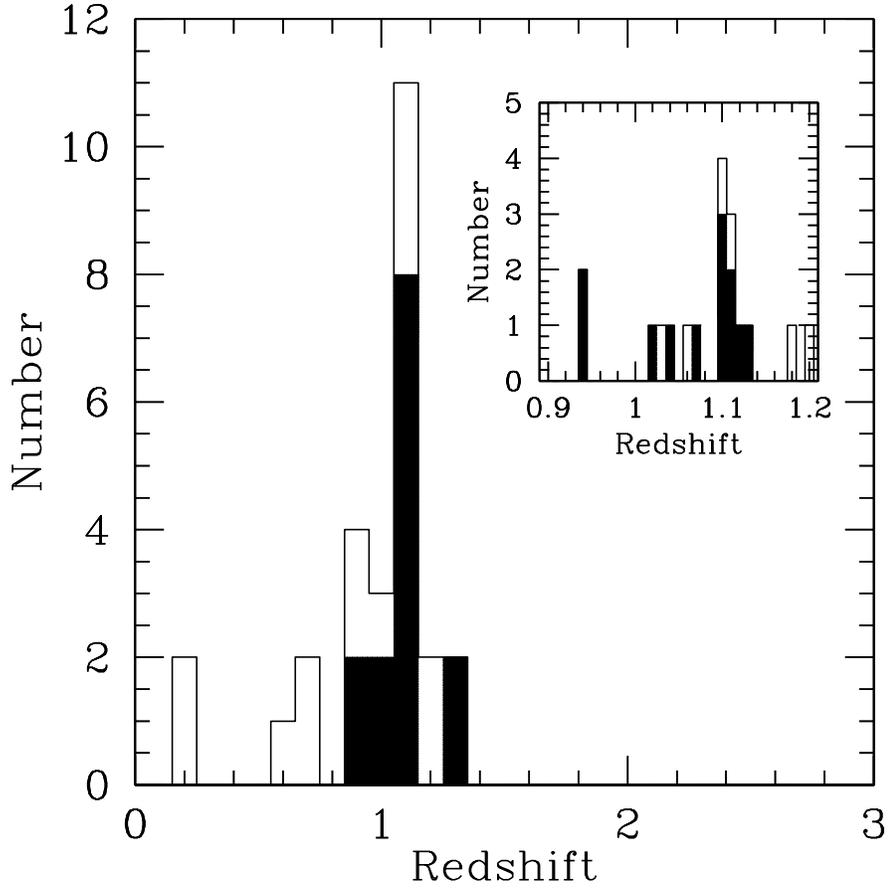}{3.8in}{0}{60}{60}{-200}{-100}
\end{center}

\caption{Redshift histogram of the 22 $K'$-detected sources in the
\mg\ field for which we have spectroscopic information.  Solid
histogram refer to sources with $I - K' > 3$.  Many of the sources with
$I - K' < 3$ were observed serendipitously, fortuitously landing on the
slitlet of a redder source.}

\label{fig_zhist}
\end{figure}

\begin{figure}[!t]
\begin{center}
\plotfiddle{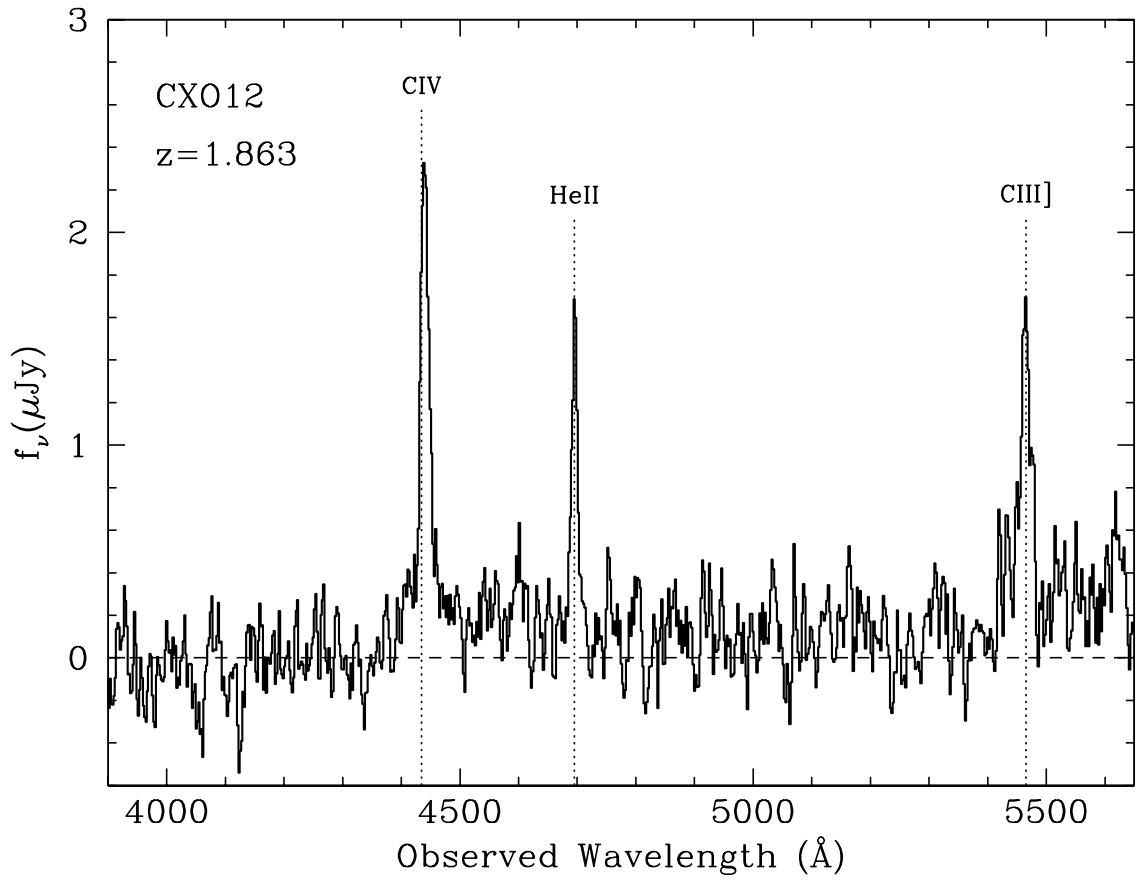}{4.0in}{-90}{60}{60}{-235}{335}
\end{center}

\caption{Optical spectrum of CXO~12, obtained with the Keck~I
telescope.  Exposure time is 5.4~ks.  The spectrum was extracted using
a 1\farcs5 $\times$ 1\farcs5 aperture and smoothed with a 15
\AA\ boxcar filter.  Prominent features are labeled.}

\label{fig_qso2}
\end{figure}

\begin{figure}[!t]
\begin{center}
\plotfiddle{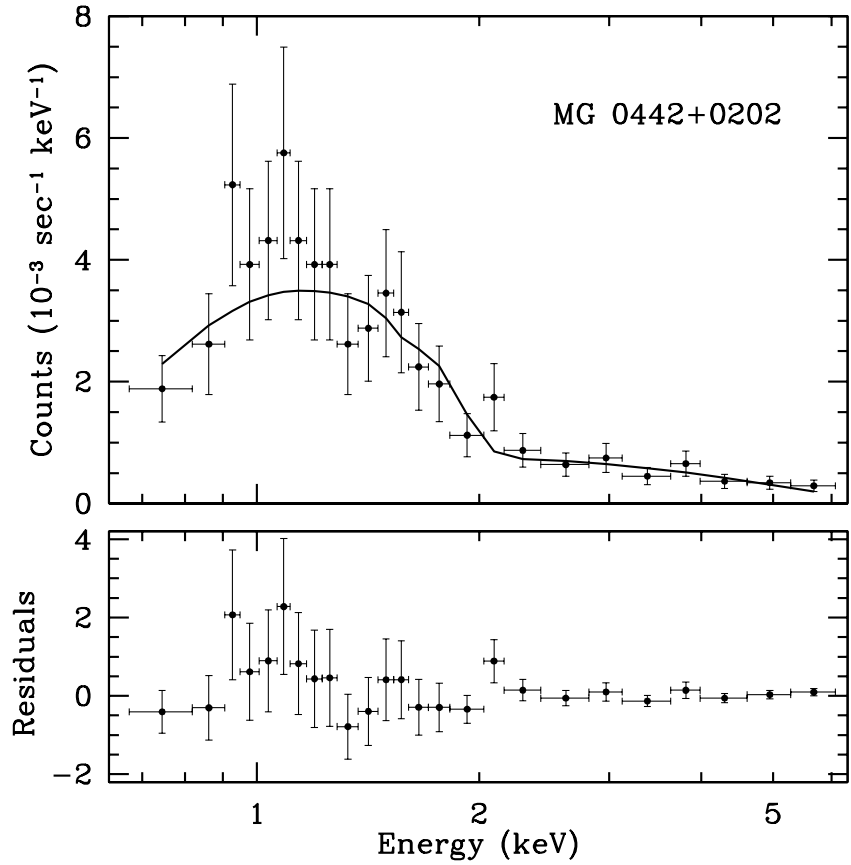}{4.2in}{0}{60}{60}{-185}{-100}
\end{center}

\caption{X-ray spectrum of \mg\ with fitted spectrum folded through the
detector response (top).  Residuals to this fit are also shown (below),
with the y-axis scale the same as in the top panel (units of $10^{-3}~
{\rm counts}~ {\rm sec}^{-1}~ {\rm keV}^{-1}$).  Each bin contains 10
counts.}

\label{fig_Xspec}
\end{figure}

\begin{figure}[!t]
\begin{center}
\plotfiddle{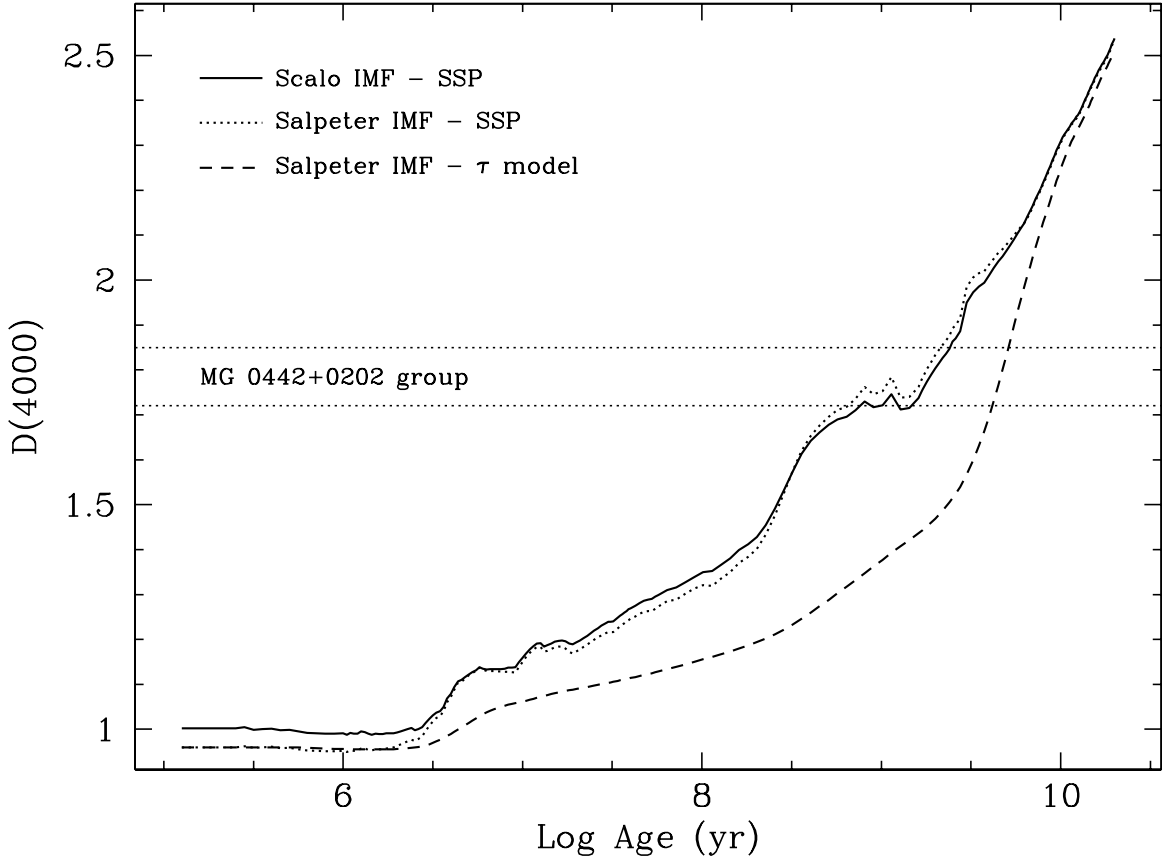}{4.0in}{-90}{60}{60}{-235}{335}
\end{center}

\caption{Amplitude of the D(4000) break, as defined by Hamilton (1985),
for the population synthesis models of Bruzual \& Charlot (1993;
version BC95).  All models assume solar metallicty and a stellar
initial mass functions within the range of $0.1 - 125 M_\odot$, using
either the Scalo (1986) or Salpeter (1955) parametrization.  SSP refers
to simple stellar populations, namely an instantaneous burst of star
formation followed by passive evolution.  The $\tau$ model refers to an
exponentially-decreasing star formation rate with a 1~Gyr e-folding
time and no gas recycling.}

\label{fig_d4000}
\end{figure}

\footnotesize
\begin{deluxetable}{lcccccccl}
\tablecaption{Summary of spectroscopic targets in \cl\ field.}
\tablehead{
\colhead{} &
\colhead{$\Delta$R.A.} &
\colhead{$\Delta$Dec.} &
\colhead{$\Delta r$} &
\colhead{$I$} &
\colhead{$K'$} &
\colhead{$I-K'$} &
\colhead{} &
\colhead{} \\
\colhead{ID} &
\colhead{(\arcsec)} &
\colhead{(\arcsec)} &
\colhead{(\arcsec)} &
\colhead{(mag)} &
\colhead{(mag)} &
\colhead{(mag)} &
\colhead{$z$} &
\colhead{Notes}}
\startdata
MG~0442+0202 & \nodata & \nodata &\nodata& 20.33 & 17.11 & 3.22 & 1.110 & radio galaxy \nl
galaxy\#25   &  $-$1.9 &  $-$2.1 &   2.9 & 22.79 & 18.79 & 4.00 & 1.105 & weak \oii \nl
galaxy\#2    &  $-$7.6 &     5.2 &   9.2 & 22.58 & 18.57 & 4.01 & 1.107 & early-type \nl
galaxy\#29   &     3.2 &    13.7 &  14.1 & 23.42 & 21.14 & 2.28 & 1.102 & early-type \nl
galaxy\#27   &     3.9 &    22.7 &  23.0 & 23.15 & 19.92 & 3.23 & 1.100 & early-type \nl
galaxy\#8    &    27.4 & $-$14.3 &  30.9 & 22.48 & 19.06 & 3.42 & 1.10  & early-type \nl
             &         &         &       &       &      &      &       & \nl
quasar\#37   &  $-$2.9 &    11.4 &  11.7 & 21.37 &$>$20.0&$<$1.37& 2.749 & quasar \nl
quasar\#43   &  $-$8.8 & $-$11.2 &  14.2 & 20.45 & 18.17 & 2.28 & 0.886 & quasar \nl
galaxy\#3    & $-$22.2 &  $-$9.5 &  24.1 & 19.98 & 17.89 & 3.09 & 0.939 & early-type \nl
galaxy\#104  &    24.3 &  $-$2.2 &  24.4 & 23.07 &$>$20.0&$<$3.07&1.060 & \oii \nl
galaxy\#101  & $-$14.5 & $-$34.7 &  37.6 & 23.17 &$>$20.0&$<$3.17&1.181 & \oii \nl
galaxy\#4    &    16.2 &    39.2 &  42.4 & 22.69 & 18.73 & 3.96 & 1.125 & early-type \nl
galaxy\#9    &    41.7 & $-$15.8 &  44.6 & 22.95 & 18.25 & 4.70 & 1.02: & early-type: \nl
galaxy\#5    &     6.1 &    51.1 &  51.5 & 22.10 & 19.01 & 3.09 & 1.133 & weak \oii \nl
galaxy\#7    &    21.4 & $-$56.6 &  60.5 & 22.75 & 18.56 & 4.19 & 0.94: & \nl
galaxy\#14   &    48.7 &    49.6 &  69.5 & 21.96 & 19.06 & 2.90 & 0.658 & \oii \nl
galaxy\#33   &    65.3 & $-$33.0 &  73.2 & 21.97 & 19.08 & 2.89 & 1.031 & \oii \nl
galaxy\#22   & $-$45.1 & $-$58.6 &  74.0 & 22.79 & 18.59 & 4.20 & 1.07: & early-type: \nl
galaxy\#17   &    55.2 &    51.6 &  75.6 & 21.13 & 18.19 & 2.94 & 0.86: & early-type: \nl
galaxy\#19   &    12.6 & $-$74.8 &  75.8 & 21.19 & 18.46 & 2.73 & 0.62  & early-type \nl
galaxy\#16   &  $-$7.4 &    79.6 &  79.9 & 18.80 & 16.77 & 2.03 & 0.194 & H$\alpha$ \nl
galaxy\#6    &    67.9 & $-$53.8 &  86.6 & 21.79 & 18.57 & 3.22 & 1.043 & early-type \nl
galaxy\#105  & $-$62.2 & $-$60.6 &  86.8 & 21.79 &\nodata&\nodata&1.120 & \oii \nl
galaxy\#13   &    39.3 & $-$77.9 &  87.3 & 22.08 & 18.52 & 3.56 & 1.261 & \oii \nl
galaxy\#106  &    38.8 & $-$81.7 &  90.4 & 20.91 &$>$20.0&$<$0.91&0.147 & H$\alpha$ \nl
galaxy\#20   &    69.7 &    58.7 &  91.1 & 20.92 & 18.26 & 2.66 & 0.693 & early-type \nl
galaxy\#102  & $-$93.2 &    10.7 &  93.8 & 20.59 &\nodata&\nodata&0.240 & \oii,H$\alpha$ \nl
galaxy\#103  &$-$103.1 &    10.5 & 103.6 & 22.82 &\nodata&\nodata&1.112 & \oii \nl
galaxy\#12   &    94.7 & $-$71.8 & 118.8 & 22.98 & 19.19 & 3.79 & 1.269 & \oii \nl
galaxy\#107  &    99.7 & $-$75.6 & 125.1 & 22.86 &$>$20.0&$<$2.86&0.193 & H$\alpha$ \nl
\enddata

\tablecomments{Astrometry is relative to the radio galaxy MG~0442+0202,
located at R.A. = 04$^h$42$^m$23.74$^s$, Dec. =
+02\deg02\arcmin19\farcs8 (J2000).  Parameters with colons indicate
uncertain measurements.  Infrared photometry is from the Lick images
for all sources except galaxies \#25 and \#29 where the IRTF photometry
is reported:  the proximity of galaxy\#25 to MG~0442+0202 precludes
accurate photometry in the poor-seeing Lick imaging, while galaxy\#29
remains undetected in the Lick imaging.  Optical photometry for sources
undetected in the $K'$ images are in 3\arcsec\ diameter apertures.}

\label{tableProp}
\end{deluxetable}
\normalsize

\footnotesize
\begin{deluxetable}{rccrrrccl}
\tablecaption{Properties of {\it Chandra} sources in CL~0442+0202 field.}
\tablehead{
\colhead{} &
\colhead{R.A.} &
\colhead{Dec.} &
\colhead{} &
\colhead{} &
\colhead{} &
\colhead{$I$} &
\colhead{} &
\colhead{} \nl
\colhead{ID} &
\colhead{(J2000)} &
\colhead{(J2000)} &
\colhead{$S^{(-16)}_{0.5-2}$} &
\colhead{$S^{(-15)}_{2-10}$} &
\colhead{${H-S}\over{H+S}$} &
\colhead{(mag)} &
\colhead{$z$} &
\colhead{Notes} }
\startdata
CXO\#2  & 04$^h$42$^m$26.90$^s$ & +02\deg00\arcmin07.4\arcsec &  20.39 &  8.78 & $-$0.30 & 23.80 & \nodata &  \nl
CXO\#4  & 04$^h$42$^m$24.06$^s$ & +02\deg02\arcmin14.2\arcsec &   5.11 & 11.91 &    0.60 & 23.5: & \nodata &  \nl
CXO\#5  & 04$^h$42$^m$23.74$^s$ & +02\deg02\arcmin19.8\arcsec & 115.35 & 51.39 & $-$0.29 & 20.91 & 1.110 & MG0442+0202 \nl
CXO\#6  & 04$^h$42$^m$23.55$^s$ & +02\deg02\arcmin31.2\arcsec &  17.90 &  3.70 & $-$0.51 & 21.37 & 2.749 & quasar\#37 \nl
CXO\#7  & 04$^h$42$^m$23.15$^s$ & +02\deg02\arcmin08.6\arcsec &  58.24 & 14.83 & $-$0.55 & 20.42 & 0.886 & quasar\#43 \nl
CXO\#8  & 04$^h$42$^m$23.17$^s$ & +02\deg02\arcmin19.8\arcsec &  44.75 & 23.08 & $-$0.22 & 22.87 & 0.835: & broad MgII: \nl
CXO\#10 & 04$^h$42$^m$24.15$^s$ & +02\deg03\arcmin11.0\arcsec &   5.91 & 10.11 &    0.37 & 22.49 & 1.133 & galaxy\#5 \nl
CXO\#11 & 04$^h$42$^m$16.36$^s$ & +02\deg02\arcmin30.6\arcsec &  27.12 & 14.13 & $-$0.18 & 22.72 & \nodata & \nl
CXO\#12 & 04$^h$42$^m$36.25$^s$ & +02\deg03\arcmin25.4\arcsec &  20.97 & 13.41 & $-$0.03 & 23.40 & 1.863 & type~II quasar \nl
CXO\#16 & 04$^h$42$^m$34.31$^s$ & +02\deg05\arcmin10.8\arcsec &  16.23 &  6.62 & $-$0.35 & 23.24 & 0.772 & \nl
CXO\#17 & 04$^h$42$^m$31.74$^s$ & +02\deg05\arcmin08.5\arcsec &  36.06 & 16.59 & $-$0.24 &$>$24.7& \nodata & \nl
CXO\#18 & 04$^h$42$^m$13.15$^s$ & +02\deg04\arcmin38.5\arcsec &  32.98 & 14.78 & $-$0.25 & 22.84 & 1.38: & broad MgII: \nl
\enddata

\tablecomments{Uncertain measurements are indicated with a colon.
$S^{(-16)}_{0.5-2}$ is the soft-band (0.5$-$2~kev) flux measured in
units of $10^{-16}~ \ergcm2s$.  $S^{(-15)}_{2-10}$ is the hard-band
(2$-$10~kev) flux measured in units of $10^{-15}~ \ergcm2s$.}

\label{tableXray}
\end{deluxetable}
\normalsize

\end{document}